# Evolution of a metastable phase with magnetic phase-coexistence phenomenon and its unusual sensitivity to magnetic-field cycling in the alloys, Tb$_{5-x}$Lu$_x$Si$_3$ ($x \leq 0.7$)

K Mukherjee, Kartik K Iyer and E V Sampathkumaran

*Tata Institute of Fundamental Research, Homi Bhabha Road, Colaba, Mumbai 400005, India*

**Abstract**
Recently, we reported an anomalous enhancement of positive magnetoresistance (*MR*) beyond a critical magnetic-field in Tb$_5$Si$_3$ in the magnetically ordered state, attributable to 'Inverse Metamagnetism'. This results in an unusual magnetic hysteresis loops in the pressurized specimens, which are relevant to the topic of 'electronic phase-separation'. In this article, we report the influence of small substitutions of Lu for Tb to see the evolution of these magnetic anomalies. We find that, at low temperatures, the high-field high-resistive phase could be partially stabilized on returning the magnetic-field to zero in many of these Lu substituted alloys, as measured by electrical resistivity (ρ). Also, the relative fractions of this phase and virgin phase appear to be controlled by a small tuning of composition and temperature. Interestingly, at 1.8 K, a sudden 'switch-over' of the value of ρ of this mixed-phase to that of the virgin phase for some compositions is observed at low-fields after a few field-cycling, indicating metastability of this mixed-phase.

1. **Introduction**
The compounds exhibiting metamagnetic transitions and 'electronic phase-separation (EPS)' have been attracting a lot of attention in condensed matter physics in recent years [1]. An intermetallic compound $Tb_5Si_3$ (with a Neél temperature of $T_N$ ~70K) and its derivatives have been recently identified by us to behave exceptionally in these respects. A strong *enhancement* of positive magnetoresistance [*MR* defined as $\{\rho(H)-\rho(0)\}/\rho(H)$] at a critical magnetic field ($H_c$), instead of a decrease of *MR* (characterizing 'metamagnetism'), is observed in these alloys [2]. This anomaly was attributed to 'Inverse metamagnetism' - a concept not so commonly known among metallic magnets – in which 'magnetic fluctuations' instead of 'ferromagnetic alignment' are favored [1, 3] beyond $H_c$. During the course of investigation of $Tb_5Si_3$ under external and chemical pressure [4-6], we made many important findings bearing relevance to the concept of EPS. For the parent compound, under ambient pressure conditions, though a strong irreversibility is observed in the plots of isothermal magnetization (*M*) and *MR versus H*, the curve while reversing the field merges with the virgin curve before attaining zero-field. However, under high pressure (e.g., 10 kbar) and for $Tb_4LuSi_3$, the high-field-phase is at least partially retained after returning the field to zero giving rise to an unusual EPS comprising of high-field high-resistive phase and low-field low-resistive phase [5, 6]. Thus, this alloy is characterized by a *MR(H)* hysteresis loop that is uncommon in magnetism, as the virgin curve lies below the envelope-loop in the plot of *MR versus H*. This unique feature provided us the basis for the present investigation of the alloy series, $Tb_{5-x}Lu_xSi_3$, for small doping of Lu ($x < 1$) to explore more carefully how such unusual *MR* loops evolve.

2. **Experimental Details**
Polycrystalline samples, $Tb_{5-x}Lu_xSi_3$ ($x$=0.1, 0.3, 0.5 and 0.7), were prepared by arc melting stoichiometric amounts of high purity (>99.9 wt.%) constituent elements in an atmosphere of high-purity argon. Single-phase nature of the specimens was ascertained by x-ray diffraction (Cu $K\alpha$). With the increase in composition of Lu, the volume of the compound as expected is found to decrease (For $x$=0.1; $a$= 8.437 Å, $c$= 6.341 Å, $V$ = 390.95 Å$^3$ and for $x$=0.7; $a$= 8.427 Å, $c$= 6.335 Å, V= 389.60 Å$^3$, with a typical error of ±0.004 Å in $a$ and $c$), establishing an increase of positive chemical pressure. The $\rho$ measurements as a function of magnetic field at 1.8 and 5 K (<120 kOe) were performed by a commercial physical property measurements system (Quantum Design) and a conducting silver paint was used for making electrical contacts of the leads with the samples. In addition, dc *M* measurements at 1.8 and 5 K (<120 kOe) with the help of a commercial vibrating sample magnetometer (Oxford Instruments) were performed for a comparative investigation.

3. **Results and Discussions**
Figure 1 (a-d) shows the magnetic-field response of *MR* (left panel) and *M* (right panel) at 1.8 K for the all the compositions for 6 cycles (cycle 1: 0-120 kOe; cycle 2: 120-0 kOe; cycle 3: 0- -120 kOe; cycle 4: -120 kOe-0 kOe; cycle 5: 0-120 kOe; cycle 6: 120-0 kOe). We have also performed ρ as a function of temperature for all the compositions to demonstrate gradual reduction in the onset of magnetic ordering temperature (from ~70 K for $x$= 0 to ~55 K for $x$= 0.7) as indicated by the temperature where ρ falls (see figure 1, inset). We have not shown the data for the parent compound here, as these were discussed in our earlier publications [2, 4] (also being similar in features to that for $x$= 0.1, presented below). From figure 1a, it is obvious that, for *x*=0.1 composition, the field where the steep rise in *MR* is observed (near 55 kOe) coincides with the field where a jump is observed in *M(H)* curve in cycle 1. Beyond this critical field, *MR* falls gradually with *H* and rises with decreasing field in cycle 2. There is a drastic fall in $\rho$ (~29 kOe) and *M* (~19 kOe) in cycle 2 implying that the reverse transition is complete within the first quadrant itself. (The critical fields inferred through these two techniques marginally differ, which we attribute to different responses to local inhomogeneities due to Lu substitution). The features observed in cycle 3 and 4 are essentially mirror images of the first quadrant. A rise as in cycle 1 and a fall similar to cycle 2 are also observed in cycles 5 and 6 respectively at almost similar fields. These features are similar to those observed for the parent compound [2, 4]. For *x*=0.3, as seen in figure 1f, the shape of the *M(H)* curve is similar to that of *x*= 0.1 showing jumps/drops at the respective cycles with a marginal decrease in the critical field values. However, in *MR(H)* (figure 1b), even though an enhancement in resistivity is observed in the first cycle (following *M(H)* behavior), no drastic fall is observed in cycle 2 while reducing the field in the first quadrant. This implies that the high-field phase is partially stabilized [7-9] at *H=0*. As argued earlier [4-6], in the high field phase, magnetic fluctuations are introduced and the magnitude of *MR* of this phase smoothly reduces (cycle 2) with *H* (essentially quadratically which is in favor of persistence of these fluctuations, as discussed in Ref. 6) as *H* is reversed towards zero. The scattering influence of this high-field phase persists in cycle 3 as well (figure 1b), but, in the cycle 4, a drastic drop in *MR* is observed near -8 kOe, implying that the virgin state can be attained after repeated field-cycling indicating metastability of the phase before cycle 5 begins. Such a drastic drop in $\rho$ has been obtained in the parent compound $Tb_5Si_3$ by applying external pressure, however, by a small application of magnetic field in the negative direction [6]. The transformation to the virgin phase is substantiated from the observation that, the curves in cycles 5 and 6 are of the same type as those in cycles 1 and 2.

Focusing on the magnetic behavior (figure 1c) of the composition *x*= 0.5, there is a subtle difference with respect to that of *x*= 0.3. In *M(H)*, the transition in cycle 1 is somewhat more broadened with a jump of a lower magnitude, as compared to previous concentrations. Also small drops/jumps are observed in cycle 2, 4 and 6 near 7, -6.2 and 5.8 kOe respectively. With respect to *MR(H)* curve of this compound, the features are similar to those for *x*=0.3 in cycles 1, 2 and 3. However, in cycle 4, while, for the composition *x*= 0.3, the high-field phase becomes unstable as one approaches zero field, for *x*= 0.5, this high field phase persists till *H*= 0 in this cycle. The nature of cycle 5 is also different from that of the other compositions (*x*<0.5), in the sense that a gradual downturn is observed with increasing *H* (for *x*= 0.5), merging with the virgin curve beyond $H_c$ only. Interestingly, in the sixth cycle, a drastic drop in $\rho$ is observed near 2 kOe, returning the $\rho$ value to that of the virgin state, implying that the 'switching' behavior is shifted to the near-end of cycle 6. If one looks at the *MR(H)* behavior of the composition *x*=0.7, no drop is observed in any cycle (other than that known for the virgin curve), and the high field phase persists in the vicinity of *H*=0 (with a monotonous decrease of magnitude of *MR* with increasing *H*) after travelling through the magnetic transition in cycle 1. The envelope curve in *MR* plot lies *above* the virgin curve like that for *x*= 1.0 [5]. Below 70 kOe, *M* decreases linearly with *H* in the reverse cycle in the first quadrant similar to that of a paramagnet, without any evidence for a first-order transition, which supports the idea of increasing dominance of 'fluctuations' in the high-field phase in zero-field after the cycle 1. *M(H)* curve however reveals a weak drop as soon as the field direction is reversed as though there is a fraction transforming to virgin phase. At this point we would like to mention that the field at which the drop is observed in *M* in cycle 2, decreases from 13 to 6 kOe from $x$=0.1 to $x$=0.5, and this is consistent with the absence of a drop in cycle 2 for $x$= 0.7.

From above discussions, it can be concluded that the high-field phase that is stabilized in zero-field coexists with the virgin phase after cycle 2 for $x > 0.1$ and, qualitatively speaking, viewed together with the *M(H)* data, the fraction of this high-field phase tends to dominate electrical conductivity with increasing *x*. To give more evidence to the idea of phase-coexistence (or electronic phase-separation), we performed MR(*H*) at a higher temperature, say at 5 K. Figure 2 (a-d) shows the MR(*H*) of the four compositions. It is seen that the signature of EPS is absent at *H*=0 (after cycling across $H_c$) for the composition *x*=0.1, as the curves falls back on virgin curves within the first quadrant in cycles 2 and 4. As we move to the concentration *x*=0.3, it appears as though only a very small fraction of high field phase dominates conductivity at *H*=0 for this temperature in cycles 2 and 4 as indicated by MR value which is closer to that of the virgin phase. For *x*=0.5, a significant fraction of the high field phase co-existing with the virgin phase dominates conductivity, as is evident from the (intermediate) value of MR after second and fourth cycles. Interestingly, for this composition, a small jump is visible close to zero field after cycles 2 and 4, as though there is a tendency to 'switching behavior' even at this temperature. The shape of the entire curve is butterfly-like with the virgin curve lying below the envelope curve. For the highest concentration *x*=0.7 in the present study, the domination of the high field phase for conductivity relative to that of the virgin phase further increases at this temperature. These trends are rather consistent with that observed at 1.8 K. Hence this figure renders clear evidence for the tuning of 'mixed-phase' and 'the dominance of electrical conductivity by a desired phase' by varying the chemical pressure.

**4. Summary**
This alloy series provides an opportunity to probe evolution of an unusual electronic phase-separation involving high-field high resistive and low-field low-resistive phase. We have shown that, under ambient pressure conditions, the electronically phase-separated magnetic phase can be tuned to change its electrical resistivity at low fields when subjected to magnetic-field cycling. There are jumps in the magnetoresistance near the vicinity of zero-field after a field-cycling for certain alloys, the origin of which is puzzling. This work thus emphasizes the need to activate theoretical understanding of various manifestations of metastable magnetic phases.

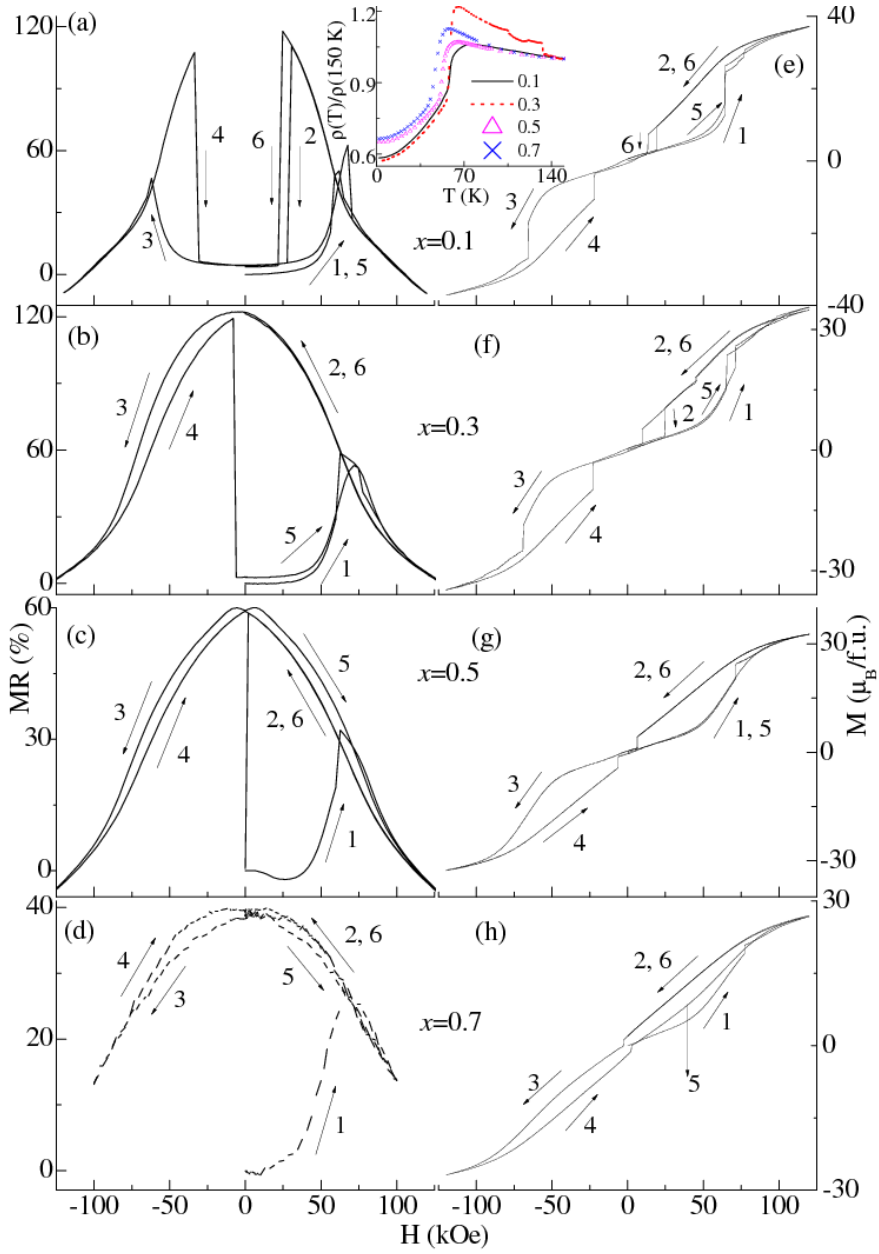

Figure 1:
(a-d) Magnetoresistance as a function of externally applied magnetic field for the sample series $Tb_{1-x}Lu_xSi_3$ with $x$=0.1, 0.3, 0.5 and 0.7 at 1.8K. (e-f) Isothermal magnetization for the same sample series at 1.8 K. Arrows and numerical are drawn as a guide to the eyes. Inset shows electrical resistivity as a function of temperature for all the alloys to highlight $x$-dependence of magnetic transition temperature.

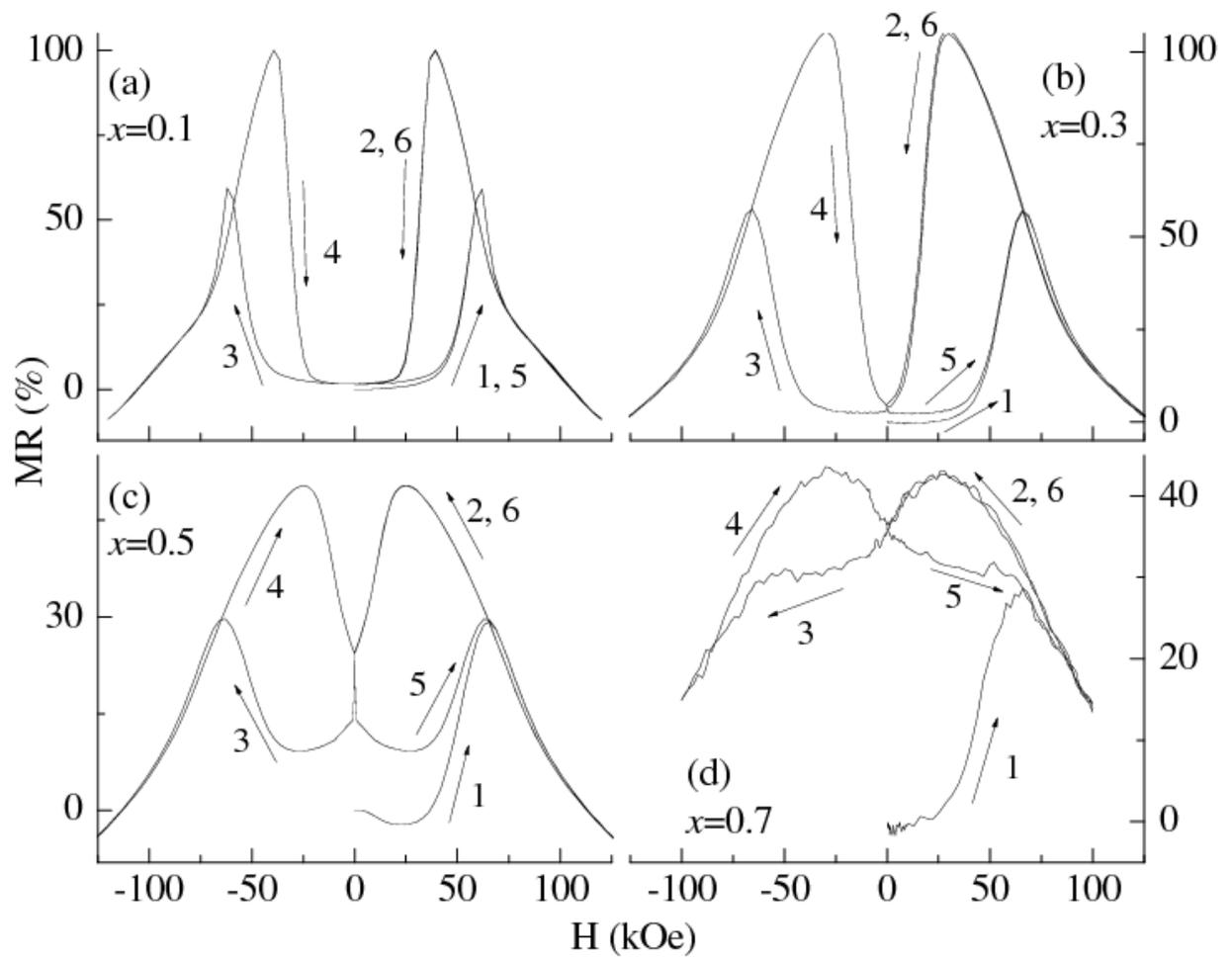

Figure 2:
(a-d) Magnetoresistance as a function of externally applied magnetic field for the sample series $Tb_{1-x}Lu_xSi_3$ with $x$=0.1, 0.3, 0.5 and 0.7 at 5 K. Arrows and numericals are drawn as a guide to the eyes.